\documentclass[a4paper,twocolumn,showpacs,amsmath,amssymb]{revtex4}
\usepackage{amsmath}
\usepackage{graphicx}

\begin{document}

\title{Spatially Separated and Correlated Atom-molecule Lasers from a Bose Condensate}

\author{Hui Jing$^{1,3}$, Wei Cai$^{1,2}$, Jing-Jun Xu$^{1,2}$ and Ming-Sheng Zhan$^{1,3}$}

\affiliation{$^1$State Key Laboratory of Magnetic Resonance and
Atomic and Molecular Physics,\\
 Wuhan Institute of Physics and Mathematics, Chinese Academe of Science, Wuhan 430071, People's Republic of China\\
 $^2$TEDA Applied Physics School, Nankai University, Tianjin,
 300457, People's Republic of China\\
 $^3$Center for Cold Atoms, Chinese Academe of Science, People's Republic of China}

\begin{abstract}
We propose a feasible scheme to create two spatially separated
atomic and molecular beams from an atomic Bose-Einstein condensate
by combining the Raman-type atom laser output and the two-color
photo-association processes. We examine the quantum dynamics and
statistical properties of the system under short-time limits,
especially the quadrature-squeezed and mode-correlated behaviors
of two output beams for different initial state of the condensate.
The possibility to generate the entangled atom-molecule lasers by
an optical technique was also discussed.
\end{abstract}


\maketitle


The experimental realizations of Bose-Einstein condensates (BEC)
in cold dilute atomic gases have provided a rich playground to
manipulate and demonstrate various properties of quantum
degenerate gases [1]. Recently rapid advances have been witnessed
for creating a quantum degenerate molecular gase via a magnetic
Feshbach resonance [2-3] or an optical photo-association (PA)
[4-5] in an atomic BEC, and the appealing physical properties of
the formed atom-molecule mixtures were investigated very
extensively under the quasi-homogeneous trapping conditions [2-5].
The technique of coherent PA not only produces a new species of
BEC but also leads to many interesting quantum statistical effects
due to its nonlinear coupling nature in the dynamics [6]. The
possible applications of a molecular condensate are expected in,
e.g., the precise matter-wave interferometry technique [7].

On the other hand, there always have been many interests in
creating an atom laser and exploring its novel properties, since
the MIT-group first realized a pulsed atom laser by using quantum
state transfer technique [8]. Successive experimental achievements
in the design and amplification of atom laser were obtained and
stimulated amounts of theoretical works in both the output
coupling and the properties of atom laser [9]. As the matter-wave
analogy of an optical laser, the atom or molecule lasers now are
expected to have some important applications in practice [7, 9].
Among the constantly expanding catalogue of atom laser schemes,
the technique of Raman output coupling was extensively studied in
order to obtain a continuous atomic beam with adjustable momentum
[10]. For a propagating atom laser, one also can further study its
manipulations such as the atomic filamentation and correlation for
a travelling beam across a magnetic Feshbach resonance region
[11].

A natural question, then, may arise: is it possible to create two
spatially separated and correlated atomic and molecular beams from
one trapped atomic condensate by combining the above two
techniques? The main purpose of this paper is to study this
interesting problem, focusing on the role of coherent two-color PA
process in the Raman atom laser coupler. We will show that, for
the classical input and control lights, the five-level
atom-molecule system can be adiabatically reduced into an
effective three-state one. And, by probing the quantum dynamics
and statistical properties of the system under short-time limits,
we can observe the quadrature-squeezed and mode-correlated
behaviors of two output beams which depends on different initial
state of the condensate. For the quantized input lights, we also
briefly discussed the possibility to generate the entangled
atom-molecule beams by applying two entangled lights.
\begin{figure}
\includegraphics[width=8cm]{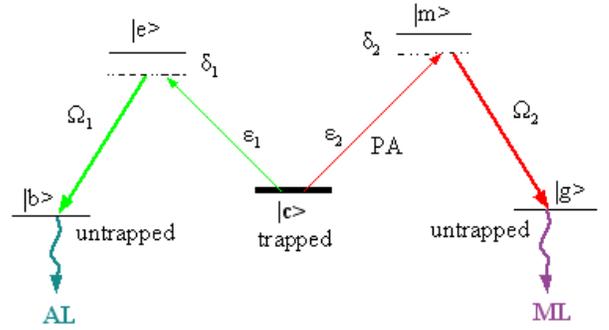}
\caption{(Color online) Illustration of creating two spatially
separated atom and molecule lasers from one trapped three-level
atomic ensemble, by combining the techniques of two-color PA and
Raman coupler. The atoms are initially trapped in the condensed
state $|c\rangle$. The atomic and the free-bound-bound transitions
are provided by two lasers, respectively: the input lights with
frequency $\varepsilon_i$ and the control lights with Rabi
frequency $\Omega_i$ ($i=1,2$). $\delta_i$ is the intermediate
detunings.} \label{1}
\end{figure}

Turning to the situation of Fig. 1, we assume for simplicity that
large number of Bose-condensed atoms are initially prepared in the
$c$ state when two strong control lights with Rabi frequency
$\Omega_i$ ($i=1,2$) are applied. The atomic coupling
$|c\rangle\rightarrow |e\rangle$ and the photo-association (PA)
$|c\rangle\rightarrow |m\rangle$ processes are described by two
input lights with frequency $\varepsilon_i$ ($i=1,2$). Note that
the two states $|e\rangle$ and $|m\rangle$ mediate the two kinds
of output coupling Raman transitions and then the stable atoms and
molecules are created in the spatially separated untrapped modes
$|b\rangle$ and $|g\rangle$, respectively. Since the atomic
collisions effects on the dynamics of the system are extensively
studied [12-13] and its strength can be tuned by the technique of
magnetic-field-induced Feshbach resonance [14-15], hence, to see
clearly the role of nonlinear PA interactions in the Raman
atom-laser output process, we ignore it for present purpose (the
strengths for the molecules or atom-molecule collisions are yet
not known [5]).

In the second quantized notation, boson annihilation operators for
the three-state atoms and the formed molecules in two states are
denoted by $\hat{c}$, $\hat{e}$, $\hat{b}$ and $\hat{m}$,
$\hat{g}$, respectively. Thereby, focusing on the different modes
couplings, the quantum dynamics of this system can be described by
the five-mode Hamiltonian ($\hbar=1$)
\begin{eqnarray}
&\hat{H}_5=&-\delta_1 \hat{e}^\dagger \hat{e}-\delta_2
\hat{m}^\dagger \hat{m}+\hat{H}_{LI}+\hat{H}_{RI},
\nonumber\\
&\hat{H}_{LI}=&\varepsilon_1(\hat{e}^\dagger\hat{c}+H.c.)+\Omega_1(\hat{b}^\dagger\hat{e}+H.c.),
\nonumber\\
&\hat{H}_{RI}=&\varepsilon_2(\hat{m}^\dagger\hat{c}\hat{c}+H.c.)+\Omega_2(\hat{g}^\dagger\hat{m}+H.c.),
\end{eqnarray}
where $\delta_i$ is the intermediate detunings and the coupling
strengths are taken as real numbers. Here, for simplicity, we have
ignored the incoherent process of the excited-state molecular
damping or the effect of molecular dissociating into those
non-condensate atomic modes [16]. In practice, the weak PA field
condition $\epsilon\ll \Omega$ could safely avoid any heating
effects in the PA process. Obviously there exists a conserved
quantity for this dynamical system: $\hat{c}^\dagger
\hat{c}+\hat{e}^\dagger \hat{e}+\hat{b}^\dagger
\hat{b}+2(\hat{m}^\dagger \hat{m}+\hat{g}^\dagger \hat{g})\equiv
N_0$, where $N_0$ is the total atom number for a condensate of all
atoms or twice the total molecule numbers.

From this five-mode Hamiltonian $\hat{H}_5$, we can write the
Heisenberg equations of motion of the excited atoms and molecules
which, by assuming $|\delta|_i$ ($i=1,2$) as the largest evolution
parameters [16] in the system or $\dot{\hat{e}}/\delta_1=0$,
$\dot{\hat{m}}/\delta_2=0$, leads to
\begin{equation}
\hat{e} \thickapprox \frac{\varepsilon_1}{\delta_1}\hat{c}+
\frac{\Omega_1}{\delta_1}\hat{b}, ~~\hat{m} \thickapprox
\frac{\varepsilon_2}{\delta_2}\hat{c}\hat{c}+
\frac{\Omega_2}{\delta_2}\hat{g},
\end{equation}
Substituting this into Eq. (1), which means adiabatic eliminations
of the excited atomic and molecular modes, yields the effective
three-mode Hamiltonian
\begin{equation}
\hat{{\mathcal
H}_3}=\lambda_1[\hat{b}^\dag\hat{c}+H.c.]+\lambda_2[\hat{g}^\dag\hat{c}\hat{c}+H.c.],
\end{equation}
where $\lambda_i=\varepsilon_i\Omega_i/\delta_i$ $(i=1,2)$. The
free motion part was ignored since it only appears as the global
phases for the output fields and has no effects on our physical
results. Also we omitted the small term of effective atomic
collisions $(\varepsilon_2/\delta_2)^2\hat{c}^{\dag 2}\hat{c}^2$
for the reasons described above. Note that, by applying the
mean-field approximation:
$(\hat{A}-\langle\hat{A}\rangle)(\hat{B}-\langle\hat{B}\rangle)\approx
0$ or $\hat{A}\hat{B}\sim
\hat{A}\langle\hat{B}\rangle-\langle\hat{A}\rangle\hat{B}$, this
three-mode Hamiltonian can be reduced into some perturbed from of
the well-known linear three-level coupling system [17-21]. The
quantum transfer technique based on the linear three-level optics
(TLO) has been intensively studied and various transfer processes
have been realized, e.g., between different internal atomic or
molecular quantum states [18], from light to atomic ensembles or
propagating atomic beams [19-20, 13, 21] and vice versa. Thereby
the mutual coherence of the two output fields can be intuitively
expected in our present system.

Now we use the short-time evolution method to analytically study
the quantum dynamics and statistical effects of this system beyond
the mean-field approximation. The Heisenberg equations of motion
for the trapped condensate, the output atomic and molecular modes
read
\begin{eqnarray}\label{4}
\dot{\hat{c}}&=&i\lambda_1 \hat{b}+2i\lambda_2 \hat{c}^\dagger
\hat{g},\nonumber\\
&\dot{\hat{b}}=&i\lambda_1 \hat{c},~~\dot{\hat{g}}=i\lambda_2
\hat{c}\hat{c},
\end{eqnarray}
respectively. Taking into account of the fact that the loss of
atoms from a condensed state occurs in impressively short time
scales (up to two hundreds of $\mu s$) [14], we now focus on the
short-time behaviors of this dynamical system by readily deriving
the solutions in second order of evolution time as
$(\delta\hat{K}(t)\equiv\hat{K}(t)-\hat{K}_0)$
\begin{eqnarray}
\delta\hat{c}(t)&=&it\lambda_1 \hat{b}_0+2it \lambda_2
\hat{c}^\dagger_0\hat{g}_0
-\frac{1}{2}t^2 \lambda_1^2 \hat{c}_0+t^2 \lambda_1 \lambda_2 \hat{b}^\dagger_0\hat{g}_0 \nonumber\\
&&-t^2 \lambda_2^2 \hat{c}^\dagger_0\hat{c}^2_0+ 2t^2 \lambda_2^2
\hat{c}_0\hat{g}^\dagger_0\hat{g}_0,
\nonumber\\
\delta\hat{b}(t)&=&it\lambda_1 \hat{c}_0-\frac{1}{2}t^2
\lambda_1^2
\hat{b}_0-t^2 \lambda_1 \lambda_2 \hat{c}^\dagger_0\hat{g}_0,\nonumber\\
\delta\hat{g}(t)&=&it \lambda_2 \hat{c}^2_0-t^2 \lambda_1
\lambda_2 \hat{b}_0\hat{c}_0 -t^2
\lambda_2^2(2\hat{c}^\dagger_0\hat{c}_0+1)\hat{g}_0.
\end{eqnarray}
Obviously, we can get a conserved quantity: $\hat{c}^\dagger
\hat{c}+\hat{b}^\dagger \hat{b}+2\hat{g}^\dagger \hat{g}\equiv
N_0$, as it should be. Note that the main feature of our present
scheme is the creation of two spatially separated atomic and
molecular beams from a three-level atomic ensemble, which is quite
different from the elegant scheme of creating two entangled atomic
beams by spin-exchange interactions from a spinor condensate [22].
In addition, comparing our reduced model to that of the simple
two-color PA process studied by, e.g., Calsamiglia et al. [16], we
note that these two three-mode models are formally similar, but
again essentially different.

Using the solutions obtained above, one can easily study the
interested quantum statistical properties for the output atoms and
molecules. As a concrete example, here we analyze the different
quantum squeezing behaviors in this system and we would see that
in the short-time limits, unlike the usual PA case, the output
molecules can exhibit an interesting squeezing-free property even
in the presence of the nonlinear atom-molecule coupling within the
propagating beam; however, if the trapped condensate is initially
prepared in a squeezed state, the output atoms and molecules may
also show the quadrature squeezed effects, indicating a possible
control of quantum statistics of the output particles by steering
the quantum state of $input$ atomic condensate.

Firstly, we assume a coherently factorized initial state of the
system, i.e., $|in\rangle =|\alpha\rangle_c|0\rangle_b|0\rangle_g$
with $\hat{c}|\alpha\rangle =|\alpha|e^{i\varphi}$. Taking into
account of the quadrature squeezed coefficients defined by [23]
\begin{equation}\label{9}
{\cal S}_i=\frac{<(\Delta \hat{{\cal
G}}_i)^2>-\frac{1}{2}|<[\hat{{\cal G}}_1,\hat{{\cal G}}_2]>}
         {\frac{1}{2}|<[\hat{{\cal
G}}_1,\hat{{\cal G}}_2]>|}, \;\; i=1,2
\end{equation}
where $\hat{{\cal G}}_{1}=\frac{1}{2}(\hat{g}+\hat{g}^{\dag})$,
$\hat{{\cal G}}_{2}=\frac{1}{2i}(\hat{g}-\hat{g}^{\dag})$, then we
can get the results for the output atomic and molecular modes
\begin{equation}\label{15}
\left (
  \begin{array}{c}
    S_{1a}(t)\\
    S_{2a}(t)
   \end{array}
\right ) = 3|\alpha|^2\lambda_1^2t^2\left (
  \begin{array}{c}
    \sin^2\varphi\\
    \cos^2\varphi
   \end{array}
\right )>0,
\end{equation}
and
\begin{equation}\label{16}
\left (
  \begin{array}{c}
    S_{1g}(t)\\
    S_{2g}(t)
   \end{array}
\right ) = |\alpha|^4\lambda_2^2t^2\left (
  \begin{array}{c}
    \sin^22\varphi\\
    \cos^22\varphi
   \end{array}
\right )>0,
\end{equation}
respectively. This means that there is still no squeezing for
these two modes even beyond the atomic undepleted approximation.
It can be readily verified that, however, for the trapped atoms
this time-independent no-squeezing effect happens only for the
case of $|\cos\varphi|=|\sin\varphi|$; for the general cases, it
really can exhibit the dynamical quadrature-squeezing behaviors.

Secondly, let us consider an initially $squeezed$ atomic
condensate (generated by, e.g., the binary Kerr-type collisions
[13]) described by [24]: $|\alpha
\rangle_s=\hat{S}(\xi)|\alpha\rangle$, where the squeezed operator
$\hat{S}(\xi)=\exp[\xi (\hat{c}^{\dag})^2-{\xi}^*\hat{c}^2]$ with
$\xi=\frac{r}{2}e^{-i\phi_s}$, as a unitary transformation on the
Glauber coherent state $|\alpha\rangle
~(\alpha\equiv|\alpha|e^{i\varphi})$. Here $r$ and $\phi_s$ denote
the squeezed strength and angle, respectively. For simplicity, we
only consider the case of squeezed vacuum state for $input$ atoms,
then for the two output beams we can get an interesting result for
the particles flux
\begin{equation}\label{9}
\langle N_g(t)\rangle_s=\eta(1+3\sinh^2r)\langle N_b(t)\rangle_s,
\end{equation}
with $\eta\equiv(\lambda_2/\lambda_1)^2$. This indicates a feature
of two-mode correlation for this system, which is somewhat similar
to the case of linearly coupling three-state system [18] or even
the scheme of spinor condensate output coupler [22]. It is clear
that the relative flux ratio is completely determined by the
coupling-strength ratio $\eta$ and the squeezed degree $r$. For
example, if we take the equal coupling strengthes or $\eta=1$, we
have $\bar{N}_g>\bar{N}_b$, i.e., the molecular beam looks
$brighter$ than the atomic beam ($r>0$) and, more $squeezing$,
much $brighter$! For possible applications, this may provide a
feasible way to measure the atomic squeezing degree of the trapped
condensate just by comparing the counted particles flux of two
output beams.

As a further illustration, we study the quantum statistical
propeties of these two output beams. It is readily to show that,
the Mandel's $Q$ parameters [24] for the output atomic and
molecular fields
\begin{equation}\label{8}
Q_{b,g}^s(\tau)\equiv\frac{\langle \Delta
\hat{N}_g^2(\tau)\rangle_s}{\langle\hat{N}_g(\tau)\rangle_s}-1 <0,
\end{equation}
which means that, at least in short-time limits, the two output
beams both satisfy the sub-Poisson distribution. It should be
remarked that the atomic depletion effect of trapped condensate
plays a key role in these results. In fact, if one makes the
Bogoliubov or undepleted approximation, the three-mode Hamiltonian
Eq. (3) becomes nothing but a trivial uncorrelated two-mode model.

In addition, we also can derive the squeezed coefficients for the
output atomic filed, i.e.
\begin{equation}\label{10}
 {\cal
 S}^{\xi_c}_{1b,2b}(t)=2\lambda_1^2t^2\sinh r(\sinh r\pm \cosh
 r\cos\phi_s),
\end{equation}
and similarly, for the output molecules we have
\begin{eqnarray}\label{11}
 \left (
  \begin{array}{c}
    {\cal
 S}^{\xi_c}_{1g}(\tau)\\
    {\cal
 S}^{\xi_c}_{2g}(\tau)
   \end{array}
\right )= \tau^2 \left [
  11(\sinh^2 r+1) \left (
  \begin{array}{c}
    \cos^2\phi_s\\
    \sin^2\phi_s
   \end{array}
   \right )-4
\right ],
\end{eqnarray}
where $\tau\equiv\lambda_2t\sinh r$. Clearly these simple results
indicate an interesting squeezed-angle-dependent squeezing effect
($r>0$): (i) for $\phi_s=2n\pi ~(n=0,1,2,...)$, we have $ {\cal
 S}_{1b}=\lambda_1^2t^2(e^{2r}-1)>0$ and $ {\cal
 S}_{2b}=\lambda_1^2t^2(e^{-2r}-1)<0$, which means that
 the quadrature component ${\cal
 S}_{2b}$ is squeezed; (ii) but for $\phi_s=(2n+1)\pi$, we have $ {\cal
 S}_{1b}={\cal
 S}_{2b}(\phi_1=2n\pi)<0$ and $ {\cal
 S}_{2b}={\cal
 S}_{1b}(\phi_1=2n\pi)>0$, i.e., the squeezing effect now
 transfers to ${\cal S}_{1b}$ component; one should note that, for
 these two cases, the output molecular field is also squeezed (both
 for the component ${\cal S}_{2g}$); (iii) for $\phi_1=(n+1/2)\pi$,
 it can be easily seen that the output atomic beam is never squeezed,
 but the molecular beam is still squeezed (but for the component
 ${\cal S}_{1g}$); (iv) but for $\phi_1=(n+1/4)\pi$, we can see that
 the molecular beam is never squeezed but the atomic beam really can
 exhibit the squeezing effect if the squeezed strength satisfies:
 $r<\ln (1+\sqrt{2})\thickapprox 0.88$. Therefore, by controlling
 the squeezed parameters of initial atomic field, one can realize
 the quantum squeezing in either of the output atomic or molecular
 beam, or in both of them!

Finally we would like to make some remarks about the mutual
coherence of the output atomic and molecular fields which can be
best characterized by the second-order cross-correlation function
\begin{equation}\label{19}
g_{bg}^{(2)}(t)= \frac{\langle \hat{b}^\dag(t) \hat{b}(t)
\hat{g}^\dag(t) \hat{g}(t)\rangle} {\langle \hat{b}^\dag(t)
\hat{b}(t)\rangle \langle \hat{g}^\dag(t) \hat{g}(t)\rangle}.
\end{equation}
It is easily verified that, however, one should write the
solutions of Eq. (4) at least in fourth order of time to determine
the results $g_{bg}^{(2)}(t)<1$ (anti-correlated states). This
means that the correlations between the two output modes are
$dynamically$ established due to the depletion of trapped atoms.
Therefore, one would ask the question about the possibility of
generating quantum correlations or entanglement between the output
atoms and molecules under the limits of large condensed atoms.

We note that, our present scheme really can realize this task if
one takes the two $input$ lights (including the
$|c\rangle\rightarrow |e\rangle$ coupling light and the PA light)
as quantized fields (denoted by $\hat{a}_{1,2}$), after the
replacement of $\hat{c},~\hat{c}^{\dagger}\rightarrow N_0$ and the
adiabatic approaches like Eq. (2), the totally seven-mode
Hamiltonian can be reduced to a simple linearly coupling four-mode
one, i.e. [13]
$$
\hat{{\mathcal
H}}_{4}=\lambda_1'[\hat{b}^\dag\hat{a}_1+H.c.]+\lambda_2'[\hat{g}^\dag\hat{a}_2+H.c.],
$$
where $\lambda_i'=\lambda_i\sqrt{N_0} ~(i=1,2)$. Obviously, this
simple form determines a factorized structure of wave function of
the system [13, 25], especially the property of perfect quantum
conversion : $\hat{a}_1\rightarrow\hat{b}$,
$\hat{a}_2\rightarrow\hat{g}$ [13]. Therefore, by preparing two
entangled input lights, one can create two spatially separated and
entangled atomic and molecular beams, which indicates a simple but
interesting optical technique for quantum control of the output
atom-molecule beams. As a concrete example, if we prepared the
input lights in a two-mode squeezed vacuum state, i.e.,
$|\zeta\rangle=exp(\zeta\hat{a}_1^\dagger\hat{a}_1^\dagger-\zeta^*\hat{a}_1\hat{a}_2)|0\rangle$,
with the squeezed parameter $\zeta=\kappa e^{-i\theta_s}$, it is
straightforward to show that, for the simplest case
$\lambda_i'=\lambda_2'=\lambda$, we have
$g_{bg}^{(2)}(t)=2+\sinh^{-2}\kappa >1$ (correlated state). This
is different form the atom-molecule correlations due to the atomic
depletions. Also, it turns out that for the two output modes:
$[g_{bg}^{(2)}(t)]^2>g_{b}^{(2)}(t)g_{g}^{(2)}(t)$, i.e., there is
violations of the Cauchy-Schwarz inequality (CSI) for them, which,
according to Reid and Walls [24], can be accompanied by violations
of Bell¡¯s inequality (nonlocality).

\bigskip

In conclusion, we have proposed a feasible scheme to create two
spatially separated and correlated atomic and molecular beams from
one trapped three-level atomic Bose-Einstein condensate, by
studying the role of two-color coherent photo-association process
in the Raman-type atom laser output coupler. We examine the
quantum dynamics and statistical properties of the output atomic
and molecular fields under short-time limits, especially the
quadrature squeezed effects and mode-correlated behaviors of the
atom-molecule beams, for different prepared states of the trapped
atomic condensate. The simple but accessible way to generate the
entangled atom-molecule lasers from a very large condensate by an
optical technique (weak input lights) was also discussed.

Of course, even more intriguing subjects can exist in the
extensive studies of this simple scheme. For example, the effects
of particles collisions in propagating modes should be considered
in further analysis although these terms can be tuned in trapped
regime by, e.g., the magnetically-induced Feshbach resonance
technique. As our earlier work on this problem within the context
of an atom laser out-coupling [13] shows, the squeezed effects
also can be predicted for the output molecules due to intrinsic
collisions. And, for the quantized input lights, especially a
quantized PA light [26], the atomic depletions effect and the
quantum noise terms should be studied by, e.g., the standard
numerical method based on $c$-number stochastic equations in
positive-$P$ representation of quantum optics [5,6]. Our
investigations here provides the possibilities for these appealing
researches.

\bigskip

\noindent This research was supported by the Wuhan Chenguang
project and NSF of China (No.10304020). H. J. would like to thank
Y. Wu and J. Cheng for useful discussions.


\begin{thebibliography}{99}

\bibitem{1}  J. R. Anglin and W. Ketterle, Nature (London) {\bf 416}, 211 (2002).
\bibitem{2}  J. Herbig, et al., Science {\bf 301}, 1510 (2003);
             E. A. Donley, et al., Nature (London) {\bf 417}, 529(2002);
             S. Inouye et al., Nature (London) {\bf 392}, 151 (1998);
             J. L. Roberts, et al., Phys. Rev. Lett. {\bf 86}, 4211 (2001).
\bibitem{3}  E. Timmermans, et al., Phys. Rep. {\bf 315}, 199 (1999);
             M. Holland, J. Park and R. Walser, Phys. Rev. Lett. {\bf 86}, 1915
             (2001).
\bibitem{4}  R. Wynar, et al., Science {\bf 287}, 1016
             (2000); N. Vanhaecke, et al.,
             Phys. Rev. Lett. {\bf 89}, 063001 (2002).
\bibitem{5}  P. D. Drummond, K. V. Kheruntsyan and H. He, Phys. Rev. Lett.
             {\bf 81}, 3055 (1998);
             J. Javanainen and M. Mackie, Phys. Rev. A {\bf 59}, R3186
             (1999);
             U. V. Poulsen and K. Molmer, Phys. Rev. A {\bf 63}, 023604
             (2001).
\bibitem{6}  J. J. Hope and M. K. Olsen,
             Phys. Rev. Lett. {\bf 86}, 3220 (2001).
\bibitem{7}  M. S. Chapman, et al., Phys. Rev. Lett. {\bf 74},
             4783 (1995);
             J. R. Abo-Shaeer, et al., Phys. Rev. Lett. {\bf 94}, 040405 (2005).
\bibitem{8}  M.-O Mewes, et al., Phys. Rev. Lett {\bf 78},
             582 (1997).
\bibitem{9}  E. W. Hagley, et al., Opt. Photon. News {\bf 13}, 22-26 (2001).
\bibitem{10} G. M. Moy, J. J. Hope and C. M. Savage, Phys. Rev. A {\bf 55}, 3631
(1997); Y. Wu and X. Yang, Phys. Rev. A {\bf 62}, 013603 (2000);
J. Ruostekoski, T. Gasenzer and D. A. W. Hutchinson, Phys. Rev. A,
{\bf 68}, 011604 (2003).
\bibitem{11} W. Zhang, C. P. Search, H. Pu, Phys. Rev. Lett. {\bf 90}, 140401 (2003).
\bibitem{12} A. S. Parkins and D. F. Walls, Phys. Rep. {\bf 303},
             1 (1998);
             Y. Shin, et al., Phys. Rev. Lett. {\bf 92}, 150401 (2004).
\bibitem{13} H. Jing, J.-L. Chen and M.-L. Ge, Phys. Rev. A {\bf 65}, 015601
(2002); {\it ibid.} {\bf 63}, 015601 (2001); C. Orzel, et al.,
Science {\bf 291}, 2386 (2001).
\bibitem{14} A. Vardi, V. A. Yurovsky and J. R. Anglin,
             Phys. Rev. A {\bf 64}, 063611 (2001);
             M. G. Moore and A. Vardi, Phys. Rev. Lett. {\bf 88}, 160402
             (2002).
\bibitem{15} C. Chin, et al., Phys. Rev. Lett. {\bf 94}, 123201 (2005).
\bibitem{16} J. Calsamiglia, M. Mackie and K.-A. Suominen,
             Phys. Rev. Lett. {\bf 87}, 160403 (2001).
\bibitem{17} K. Eckert, et al., Phys. Rev. A {\bf 70}, 023606 (2004);
             A. D. Greentree, et al., Phys. Rev. B {\bf 70}, 235317
             (2004).
\bibitem{18} K. Bergmann, H. Theuer and B. W. Shore, Rev. Mod. Phys. {\bf 70}, 1003 (1998).
\bibitem{19} M. Fleischhauer and M. D. Lukin, Phys. Rev. Lett. {\bf 84}, 5094 (2000).
\bibitem{20} U. V. Poulsen and K. Molmer, Phys. Rev. Lett. {\bf 87}, 123601 (2001);
             M. Fleischhauer and S. Q. Gong, Phys. Rev. Lett. {\bf 88}, 070404 (2002).
\bibitem{21} C. P. Search, Phys. Rev. A {\bf 64}, 053606 (2001);
             S. A. Haine and J. J. Hope, Phys. Rev. A {\bf 72}, 033601.
             (2005).
\bibitem{22} H. Pu and P. Meystre, Phys. Rev. Lett. {\bf 85}, 3987 (2000);
             L.-M. Duan, et al.,
             Phys. Rev. Lett. {\bf 85}, 3991 (2000).
\bibitem{23} V. Buzek, A. V. Barranco and P. L. Knight, Phys. Rev. A
             {\bf 45}, 6570 (1992).
\bibitem{24} D. F. Walls and G. J. Milburn, {\it Quantum Optics},
             (Springer-Verlag, Berlin, 1994).
\bibitem{25} C. Sun, et al. Commun. Theor. Phys. {\bf 29},
             161 (1998).
\bibitem{26} H. Jing and M. Zhan, e-print quant-ph/0512149
(2005).


\end{thebibliography}
\end{document}